\documentclass[twocolumn,showpacs,amsmath,amssymb,pra]{revtex4}
\usepackage{graphicx}% Include figure files
\usepackage{dcolumn}% Align table columns on decimal point
\usepackage{bm}% bold math
\usepackage[latin1]{inputenc}
\usepackage{float}

\begin{document}

\title{Two electron entanglement enhancement by an inelastic scattering process}

\author{Alexander L\'opez}
\affiliation{Centro de F\'isica, Instituto Venezolano de Investigaciones Cient\'ificas.
IVIC, Apartado 21827,Caracas 1020 A, Venezuela.}

\author{V\'ictor M. Villalba}
\affiliation{Centro de F\'isica, Instituto Venezolano de Investigaciones Cient\'ificas.
IVIC, Apartado 21827,Caracas 1020 A, Venezuela.}

\author{Ernesto Medina\footnote{corresponding author, ernesto@ivic.ve}}
\affiliation{Centro de F\'isica, Instituto Venezolano de Investigaciones Cient\'ificas.
IVIC, Apartado 21827,Caracas 1020 A, Venezuela.}
\date{\today}

\begin{abstract}
In order to assess inelastic effects on two fermion entanglement production, we address an exactly solvable two-particle scattering problem where the target is an excitable scatterer. Useful entanglement, as measured by the two particle concurrence, is obtained from post-selection of oppositely scattered particle states. The $S$ matrix formalism is generalized in order to address non-unitary evolution in the propagating channels. We find the striking result that inelasticity can actually increase concurrence as compared to the elastic case by increasing the uncertainty of the single particle subspace. Concurrence zeros are controlled by either single particle resonance energies or total reflection conditions that ascertain precisely one of the electron states. Concurrence minima also occur and are controlled by entangled resonance situations were the electron becomes entangled with the scatterer, and thus does not give up full information of its state.  In this model, exciting the scatterer can never fully destroy phase coherence due to an intrinsic limit to the probability of inelastic events.
\end{abstract}
\maketitle

\section{introduction}

The study of mechanisms for generating entanglement in the solid state environment is an interesting and active research field \cite{1,2} due to the relevance of entanglement as a resource for quantum information processing \cite{3}. Among several proposals to generate entanglement one can consider the orbital or internal degrees of freedom of either bosonic or fermionic systems. Some approaches involve direct Coulomb interactions in quantum dots \cite{4} and interference effects \cite{5}, phonon mediated interactions in superconductors \cite{6} and Kondo-like scattering of conduction electrons \cite{7}. One can also analyze the production of entanglement in scattering processes with either interacting or independent electrons \cite{8}. The latter case is of interest because the non local character of the quantum correlations (entanglement) emerges as a consequence of the scattering process. Usually, one restricts the analysis to closed systems where the scattering is elastic. However, in most experimental settings the system under study is coupled to external degrees of freedom and therefore renders the system open to interaction with `reservoirs'. 

In order to understand the role of inelasticity, in this work we address the problem of entanglement generation for two noninteracting electrons scattered independently by a point scatterer with internal structure, focusing on the role of bound states as well as inelastic scattering in entanglement production. One can achieve both of these features in a scattering process by an atom or molecule that has at least one ground state and an excited state for its internal structure. Overcoming a critical energy for the incoming electron opens up the excited state of the molecule, absorbing part of the incoming energy, making the process for the ground state channel inelastic. The latter is quantified by a loss of unitarity in the ground state channel. Nevertheless, the model as a whole is unitary so that the probability is leaking through the newly opened excited state, with the interesting feature that now we have orthogonal non interfering amplitudes in orthogonal channels. Such a decoherence mechanism has been discussed extensively by Foa et al \cite{10} in the context of electron-phonon coupling.  As pointed out in that reference, this is not the same as the action of a quasi-elastic voltage probe that randomizes the phases but does not destroy unitarity. Our approach is more related to inelastic voltage probe effects in which fictitious voltage probes are attached to the system such that there is an exchange of particles between the system and the reservoirs. Particles escaping from the conductor, enter the reservoir. These outgoing particles are substituted by particles coming from the reservoirs whose energies and phases are uncorrelated with those of the escaping electrons \cite{ButtikerReview}. The effects of voltage probes on entangled states has recently been addressed by Prada, Taddei and Fazio \cite{Prada} and also in reference \cite{Beenakkerymas}. In reference \cite{Prada2}, the authors explicitly deal with the inelastic case and focus on inelastic voltage probe effects in entanglement detection and discrimination. A related dephasing model was first addressed in reference \cite{8}.

We emphasize that the situation described above is not equivalent to the simple two channel case since the scattering entity absorbs energy. This is a vital difference since it results in the incoherent sum of the transmissions in each channel\cite{10}. An important feature of the one electron problem is that it is exactly solvable so that there are no pitfalls regarding the possible output entangled states. A perturbation approximation or disregarding part of the wavefunction, arguing higher orders in transmission or reflection amplitudes can result in misleading conclusions as pointed out in reference \cite{11}. Entanglement is indeed a very subtle global feature in which perturbation theory has to be considered carefully \cite{5,11}.
%-----------------------begin------------------Fig.1
\begin{figure}
\includegraphics[width=8.5cm]{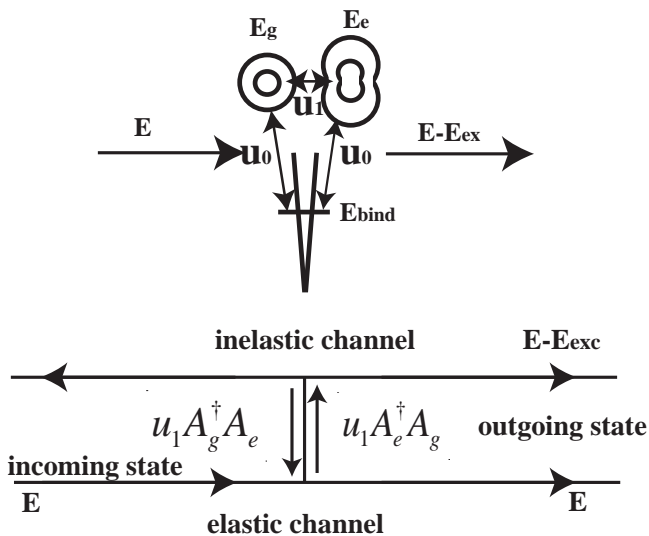}
\caption{\label{fig1} {Top panel: Model inelastic scattering system described by the Hamiltonian. The elastic scattering strength $u_0$ is dependent on the occupation of the coupled excitable scatterer, while inelastic scattering is measured through coupling $u_1$. Once a threshold is reached for the incoming particle, giving the scatterer an excitation energy  $E_{exc}=E_e-E_g$, real processes can occur in which the scatterer is excited and the scattered particle  reduces its energy to $E-E_{exc}$. $E_g$ ($E_e$) corresponds to the scatterer's ground (excited) state energy and $E_{bind}$ is the only state supported by the bare delta potential.  Bottom panel: There are two channels for scattering coupled by inelastic effects. The two outgoing channels are orthogonal and thus cannot interfere as it is the probabilities that add and not the amplitudes\cite{10}.}}
\end{figure}
%----------------------end---------------------Fig.1
The main goal of our work is to determine how the inelastic nature of the scattering process modifies the wave function of the fermion system changing the mutual information encoded in the entanglement among the particles. In order to quantify entanglement we use the concurrence \cite{12} and take advantage of Beenakker's second quantized scheme \cite{8} which takes full account of the scattering ingredients for two non interacting fermions in a second quantized description. We model a one dimensional scattering problem in which electrons are scattered independently. For simplicity and for the sake of an exact solution, we take the scatterer as a delta potential with an internal two level structure that can model the scattering of particles by an atom or molecule with the possibility of inelastic processes due to the excitation of the scatterer. Here, we determine that such real inelastic processes are relevant to entanglement generation as compared to elastic ones for which it has been found that electron correlations can, in principle, be understood in terms of virtual processes \cite{14}.  
\section{The model}
\begin{figure}
\includegraphics[width=8cm]{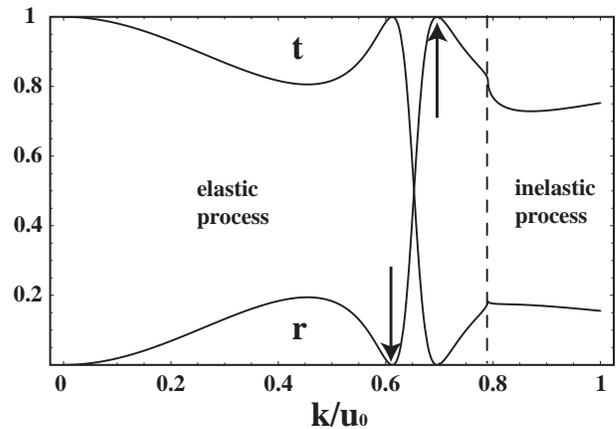}
\caption{\label{fig2} {The reflection and transmission coefficients as a function of the incoming momentum. For the elastic regime $|r|^2+|t|^2=1$, while for $k$ beyond the energy $E_e-E_g$ the process is inelastic.The arrows indicate resonance and totally reflective conditions.}}
\end{figure}

The set up for the model considered is depicted in Fig.\ref{fig1} where a two electron product wave function is injected into the potential domain which consists of an attractive delta  potential at the origin holding two possible energy states; the ground state with energy $E_g$ and an excited state with energy $E_e$. The total Hamiltonian for the system $H=H_0+V$ consists of a free part $H_0$ and the interaction $V$ given by
\begin{eqnarray}
H_0&=&\sum_k \varepsilon_k a^{\dagger}_k a_k+E_gA^{\dagger}_{g}A_{g}+E_eA^{\dagger}_{e}A_{e},\\
\nonumber
V&=&-\frac{\hbar^2}{2m}\delta(x)[u_0(A^{\dagger}_{g}A_{g}+A^{\dagger}_{e}A_{e})+u_1(A^{\dagger}_{g}A_{e}+A^{\dagger}_{e}A_{g})].
\end{eqnarray}
Here, $a^{\dagger}_k$ ($a_k$) creates, (annihilates) a particle of momentum $k$ with energy  $\varepsilon_k$. On the other hand, the operator $A^{\dagger}_g$ ($A^{\dagger}_e$) creates the scatterer in the ground (excited) state, while $u_0$ is the strength of the delta potential and $u_1$ is a coupling allowing transitions between the ground state and the excited state. Both these parameters are positive. This simple model was introduced by Lipkin \cite{Lipkin} in the context of inelastic scattering by a nucleus. Since the electrons do not interact, one first analyzes the single particle scattering process. The wave function describing either electron scattered off the potential can be written in the form
\begin{equation}\label{psi}
|\Psi\rangle=\psi_g(x) A^{\dagger}_{g}|0\rangle+\psi_e(x) A^{\dagger}_{e}|0\rangle,
\end{equation}
where $\psi_g(x),\psi_e(x)$ describe the spatial electronic states. We use the standard procedure for the treatment of delta potentials. The discontinuity of the derivative gives the relation
\begin{eqnarray*}
\lim_{\varepsilon\to 0^{+}} \left[ -\frac{\hbar^2}{2m}\frac{d\Psi(x)}{dx}\right]^{+\varepsilon}_{-\varepsilon}-\frac{\hbar^2}{2m}u_o\{A^{\dagger}_{g}A_{g}+A^{\dagger}_{e}A_{e}\}\\ -\frac{\hbar^2}{2m}u_1\{A^{\dagger}_{g}A_{e}+A^{\dagger}_{e}A_{g}\}\Psi(0)=0. 
\end{eqnarray*}
Hence, by using the wavefunction (\ref{psi}), along with the orthogonality relations for the scatterer states, one gets the following system of coupled equations
\begin{eqnarray}
\frac{d\psi_g(0^{+})}{dx}-\frac{d\psi_g(0^{-})}{dx}+u_0\psi_g(0)+u_1\psi_e(0)=0, \label{boundary1}\\
\frac{d\psi_e(0^{+})}{dx}-\frac{d\psi_e(0^{-})}{dx}+u_0\psi_e(0)+u_1\psi_g(0)=0.\label{boundary2}
\end{eqnarray}
Since only the even parity eigensolution is scattered by this potential, the phase shift $\delta_1$ associated with the odd parity eigenstate vanishes identically i.e. the odd parity solution is zero where the potential is finite, thus no phase shift ensues. Inclusion of the asymmetric potential case gives an additional phase shift (inversion asymmetry). We will not deal with this situation here since the results do not change qualitatively. Following Lipkin, we consider a scattering problem where the scatterer is initially in its ground state and choose $\psi_g(x)$ as a standing wave solution with a phase shift $\delta_0$ and $\psi_e(x)$ having only outgoing waves
\begin{eqnarray}
\psi_g(x)&=&\alpha\cos(k|x|+\delta_0),\nonumber\\
\psi_e(x)&=&\beta e^{ik_e|x|}\nonumber,
\end{eqnarray}
with
\begin{eqnarray}
k^2&=&\frac{2m(E-E_g)}{\hbar^2},\nonumber\\
k_{e}^2&=&\frac{2m(E-E_e)}{\hbar^2}=k^2-\frac{2m(E_e-E_g)}{\hbar^2}.\label{excited}
\end{eqnarray}
This selection of boundary conditions is appropriate since de-excitation of the scatterer leads to an outgoing free particle. Using equations (\ref{boundary1}) and (\ref{boundary2}) and solving for the phase shift, one finds 
\begin{equation}
\label{phase}
\tan\delta_0=\frac{u_0}{2k}\left(1-\frac{u_{1}^2}{u_0^{2}+4k_e^2}\right)+\frac{ik_eu_{1}^2}{k(u_0^{2}+4k_e^{2})}.
\end{equation}
The complex nature of the phase shift properly describes the relevant features of the inelastic scattering. With this formulation the scatterer is described, implicitly, by the $k$ dependences of the phase shift or of the reflection and transmission amplitudes.  This is nicely illustrated by considering the intensity of the reflected and transmitted waves as given by
\begin{equation}
\label{intensity}
|t|^2+|r|^2=\frac{e^{-4{\rm Im}\delta_0}+1}{2}\leq 1 
\end{equation}
where the transmission $t$ and reflection $r$ scattering amplitudes are defined in terms of the phase shifts as
\begin{eqnarray}
\label{tras}
t&=&\frac{\left(e^{2i\delta_0}+1\right)}{2},\\
r&=&\frac{\left(e^{2i\delta_0}-1\right)}{2}.
\label{refle}
\end{eqnarray}
When the electron energy $E$ is lower than the energy of the excited state $E_e$, there is no inelastic scattering and the equality is satisfied for Eq.\ref{intensity} (see Fig.\ref{fig2}). In this case, $k_e$ as defined in (\ref{excited}) is purely imaginary giving a real phase shift $\delta_0$ and there are no losses. However, when the electron energy is high enough to excite the atom, $k_e$ is purely real giving a positive imaginary part for the phase shift and the intensity (\ref{intensity}) is lower than one.
%-----------------------begin------------------Fig.2
%----------------------end---------------------Fig.2

When one considers the poles of the $S$ matrix, which occur when $\tan\delta_0=-i$, and following the discussion in \cite{9}, we find for the elastic case ($u_1=0$) that bound states exist for both the excited and ground state momenta satisfying $k_e, k=u_0/2i$. On the other hand, if the inelastic couping $u_1$ is finite, two possible situations emerge depending on whether the binding energy ($E_{binding}=-\hbar^2u_0^2/8m$) of the attractive delta potential, is i) larger or ii) smaller than the excitation energy $E_{exc}=E_e-E_g$ of the scatterer. In the first case, the two bound states are stable while for the latter situation, the bound state of the particle with the excited state of the scatterer can decay to the ground state of the scatterer plus a free particle. In short, two stable bound states, in the elastic case, turn into one stable and one unstable one above the inelastic threshold. The explicit description of the scatterer is completely contained in the behavior of the amplitudes $r$ and $t$, and thus we can now directly treat the one electron problem as a simple scattering situation.

\section{Two particle scattering}
Having considered the one-particle case in detail we now determine the role of inelastic scattering in two electron entanglement production. Since the particles do not interact, we can construct the $S$ matrix for two electron scattering from the one particle $S$ using the matrix approach of Beenakker et al\cite{8}. The two fermion incoming product state is written as $|\Psi_{in}\rangle =a^{\dagger}_{in,k_1}(\epsilon)a^{\dagger}_{in,k_2}(\epsilon)|0\rangle$, where $\epsilon$ is the one particle energy, and the subscript ${in,i}$ denotes incoming on channel $i$. One can rewrite this product state in the space of all incoming and outgoing channels as
\begin{displaymath}
|\Psi_{in}\rangle = {a_{in}^{\dagger} \choose  b_{in}^{\dagger}}^T
\left (\begin{array}{cc}
\frac{i}{2}\sigma_y & 0\\
0 & 0
\end{array}\right )
{a_{in}^{\dagger} \choose b_{in}^{\dagger}}|0\rangle,
\end{displaymath}
%More generally
%\[\begin{pmatrix}2 & 3 \\ 1 & 17\end{pmatrix}\]
the annihilation operators (2-vectors) $a_{in}$ and $b_{in}$ destroy left and right incoming particles respectively. The intermediate $4\times 4$ matrix (we shall denote $\Sigma$) contains $\sigma_y$, the corresponding $2\times 2$ Pauli matrix. The one particle input-output amplitude relation is given by the $S$ matrix through
\begin{equation}
{a_{out}\choose b_{out}}=S {a_{in}\choose b_{in}},
\end{equation}
since the process is in general nonunitary, we must consider $S\neq (S^{\dagger})^{-1}$, so the output state is given by the expression
\begin{displaymath}
|\Psi_{out}\rangle = {a^{\dagger}_{out}\choose b^{\dagger}_{out}}^T\mathcal{S} \Sigma \mathcal{S}^T {a^{\dagger}_{out}\choose b^{\dagger}_{out}}|0\rangle.
\end{displaymath} 
where $\mathcal{S}=(S^{\dagger})^{-1}$ and superindex $T$ denotes the transpose. Clearly,  ${\rm \bf W}=\mathcal{S}\Sigma\mathcal{S}^T$ is antisymmetric. Since the $S$ matrix has the regular partition in $2\times 2$ blocks, 
\begin{equation*}
S=\left (\begin{array}{cc}
r & t' \\
t & r'
\end{array}\right ),
\end{equation*}
where $r,r',t,t'$ are $2\times 2$ matrices, the $\mathcal{S}$ matrix will also have the same structure
\begin{equation}
\label{maD}
\mathcal{S}=\left (\begin{array}{cc}
\mathcal{R} & \mathcal{T}' \\
\mathcal{T} & \mathcal{R}'
\end{array}\right ),
\end{equation}
where
\begin{eqnarray*}
\mathcal{R}&=& [r^{\dagger}-t^{\dagger}(r'^{\dagger})^{-1}t'^{\dagger}]^{-1},\\
\mathcal{T}'&=&-\mathcal{R}t^{\dagger}(r'^{\dagger})^{-1},\\
\mathcal{T}&=&-(r'^{\dagger})^{-1}t'^{\dagger}\mathcal{R},\\
\mathcal{R}'&=&(r'^{\dagger})^{-1}-(r'^{\dagger})^{-1}t'^{\dagger}\mathcal{T}'.
\end{eqnarray*}
While it is obvious that the $\mathcal{S}$ matrix has the structure of Eq.\ref{maD} when the reflection and transmission matrices are diagonal (no channel mixing), such structure can be shown for the general case. 

One can compute the concurrence for a bipartite system of identical fermions in the second quantized form, following the general expression of Schliemann, Loss and MacDonald\cite{LossConcurrence}. This is done identifying the ${\rm \bf W} $ matrix in the expansion $|\Psi_{out}\rangle =\sum_{\alpha,\beta}{\rm W}_{\alpha\beta}~c^{\dagger}_\alpha c^{\dagger}_\beta|0\rangle$,
where $\alpha,\beta\in \{1,2,3,4\}$ \cite{LossConcurrence}. We have made the identification $(c^{\dagger}_1,c^{\dagger}_2,c^{\dagger}_3,c^{\dagger}_4)= (a_{out}^{\dagger},b_{out}^{\dagger})$. One can define the dual state as $| {\tilde \Psi_{out}}\rangle=\sum_{\alpha,\beta}{\rm {\widetilde W}}_{\alpha\beta}~c^{\dagger}_\alpha c^{\dagger}_\beta|0\rangle$, where ${\rm {\widetilde W}}_{\alpha\beta}=1/2\sum_{\gamma,\delta}\varepsilon^{\alpha\beta\gamma\delta}{\rm W}^*_{\gamma,\delta}$, $\varepsilon^{\alpha\beta\mu\nu}$ is the totally antisymmetric unit tensor in 4 dimensions and ${\rm \bf W}^*$  is the complex conjugate. The expression for the concurrence\cite{LossConcurrence} in terms of the previous definitions is then $\eta=|\langle {\tilde \Psi_{out}}|\Psi_{out}\rangle|=\varepsilon^{\alpha\beta\mu\nu}{\rm W}_{\alpha\beta}{\rm W}_{\mu\nu}$. In terms of the {\bf W} matrix elements, $\eta=8|{\rm W}_{12}{\rm W}_{34}+{\rm W}_{13}{\rm W}_{42}+{\rm W}_{14}{\rm W}_{23}|$. For the full output wavefunction the concurrence is easily shown to vanish identically no matter the structure of the potential whether it be inversion asymmetric or inelastic. One recovers the fully elastic case fully by setting $\mathcal{S}=S$ (see ref. \ref{maD}) as expected, for unitary processes. 

If one postselects the output state so that one only detects states where one particle is reflected and the other transmitted (simultaneous detection on opposite sides of the barrier), one arrives at the appropriately normalized wavefunction
\begin{equation}
|\Psi_{PS}\rangle=\frac{1}{\sqrt{{\rm Tr}\gamma\gamma^{\dagger}}}a_{out}^{\dagger}\gamma b_{out}^{\dagger}|0\rangle,
\end{equation}
where $\gamma=\mathcal{R}\sigma_y\mathcal{T}^{T}$. From this form, and restricting ourselves to the non-channel mixing scenario, one can readily compute the concurrence to be 
\begin{equation}\label{eta}
\eta=\frac{2|\mathcal{R}_{22}||\mathcal{T}_{11}||\mathcal{R}_{11}||\mathcal{T}_{22}|}{|\mathcal{R}_{22}|^2|\mathcal{T}_{11}|^2+|\mathcal{R}_{11}|^2|\mathcal{T}_{22}|^2},
\end{equation}
where $\mathcal{R}_{ii}$ and $\mathcal{T}_{ii}$ ($i=1,2$) are written in terms of the reflection and transmission matrix as given in the following expressions
\begin{eqnarray*}
\mathcal{R}_{ii}^{*}&=&\left(\frac{r_{ii}}{r_{ii}^2-t_{ii}^2}\right), \\
\mathcal{T}_{ii}^{*}&=&\left(\frac{-t_{ii}}{r_{ii}^2-t_{ii}^2}\right) ,
\end{eqnarray*} 
where $*$ denotes complex conjugation. As expected, for unitary process $\mathcal{R}_{ij}=r_{ij}$, $\mathcal{T}_{ij}=t_{ij}$. Channel mixing does not change the scenario qualitatively, and one always needs to post-select in order to find a finite concurrence. It does not make sense to neglect, according to perturbative arguments, terms higher order in the transmission or reflection amplitudes, since entanglement is a global property. Any disregard for small terms (for example, if $t<<1$, one cannot disregard terms of order $t^2$)  amounts to a projection of the system that may artificially generate or destroy entropy of the bipartite system. 

A physically motivated way to quantify `useful' entanglement in this problem was suggested by Wiseman and Vaccaro \cite{WisemanVaccaro} (see also \cite{BeenakkerReview}) where Fock space is separated in sectors conserving particle number. In our case  there are three sectors; If we have a particle detector A left and B right of the scatterer, the sectors will be two particles in detector A (two particles reflected), two particles in detector B (two particles transmitted) and one particle in each detector. Then one adds the separate concurrences of the projected density matrices (for each sector) eliminating coherences between states of different particle number. This achieves the same result as our post-selection.

\section{results}

We now analyze the two particle scattering scenario including inelastic effects. We first note that  there are some natural units to express energy and momentum and the strength of inelasticity in the problem. The parameter $u_0$, the strength of the delta potential, has units of momentum, and the binding energy of the same delta potential $E_{bind}=\hbar^2u_0^2/8m$ we adopt as an energy unit. The ratio $g=u_1/u_0$ is a natural measure for inelastic coupling. 

\begin{figure}
\includegraphics[width=8cm]{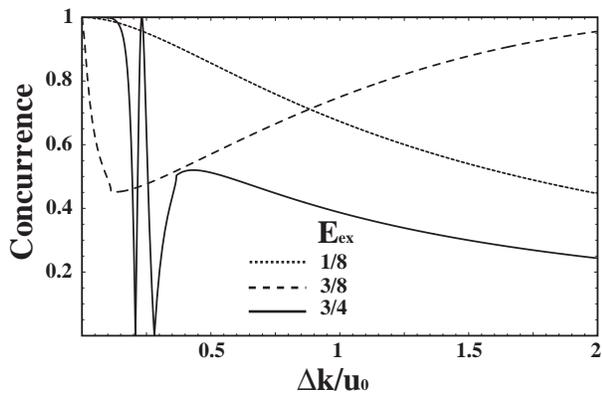}
\caption{\label{fig3} {The concurrence as a function  of the incoming particle momenta for fixed $k_1=u_0/2$ and inelastic strength $g=u_1/u_0=1/2$. i) For the lowest excitation energy of the scatterer, both momenta are always above the inelastic threshold momentum $\sqrt{E_{exc}}$ (in units of $u_0$) and the concurrence decays monotonically as a function of the particle momentum difference (dotted line). ii) When $E_{exc}$ increases one of the particle momentum is always below the inelastic threshold. A minimum concurrence is observed when the second particle momentum hits the inelastic threshold (dashed line) as $\Delta k$ increases.  iii) Both particle momenta are below the inelastic threshold (continuos line) within the $\Delta k$ range depicted, and the concurrence has two exact zeros due to corresponding zeros in the one particle reflection and transmission amplitudes (see arrows in Fig.\ref{fig2}). }}
\end{figure}

For any fixed momentum for one of the two incoming particles, we distinguish three possible situations: i) Both particles are below the threshold ii) one of them is always below and the other goes from below to above and iii) both momenta are always above the inelastic energy threshold. The concurrence, for the three cases above, is depicted in Fig.\ref{fig3} for one fixed incoming momentum $k_1=u_0/2$ as a function of the two-particle momentum difference $\Delta k/u_0$ and for three values of the normalized internal excitation energy $E_{exc}=(E_e-E_g)/(4E_{bind})$ (thus $E_{exc}$ values are shown dimensionless). The strength of inelastic effects has been set to $g=u_1/u_0=1/2$. The three different cases discussed above are represented by: i) continuous ii) dashed and iii) dotted line, respectively.
%-----------------------begin------------------Fig.3
%----------------------end---------------------Fig.3
Fig. \ref{fig3} shows that for the highest excitation energy one has two minima for the concurrence that correspond to first, a zero in the one particle transmission amplitude and second, a zero in the reflected transmission amplitude. Such zeros of the concurrence are obvious consequence of the reduction of the post-selected state to a product state. The origin of the entanglement, for the post selected state, is the uncertainty regarding which particle gets transmitted and which reflected. This creates the quantum correlation that if one particle is detected at transmission the other must be reflected. If a certainty of transmission of one of the states arises, the entanglement  is broken; {\it The resonance measures the state}. 

It is useful to note that in the previous situation (only elastic scattering) we have a product state between any one of the electrons and the scatterer, since the latter is definitely in the ground as it cannot be excited. In a sense the only uncertainties in the scattering process are due to the reflection and transmission process. As the excitation energy is increased and one of the particle's momentum is set below the threshold (dashed line inf Fig.\ref{fig3}, for the whole $\Delta k$ region) a minimum of the concurrence develops as the second momentum hits the threshold. This minimum corresponds to the onset of inelastic processes, opening the excited channel for the scatterer. Now the scattered particle and the scatterer become entangled i.e. whenever the electron is in state $\psi_g(x)$ the scatterer is in $A_g^{\dagger}|0\rangle$, while if the electron is in $\psi_e(x)$ the scatterer is in state $A_e^{\dagger}|0\rangle$ (see Eq.\ref{psi}). Hitting the threshold for inelasticity from below can only determine the state of the inelastically scattered electron thus reducing the concurrence. Nevertheless, in contrast to a standard resonance, discussed before, where the state of the scattered electron is completely determined, the inelastic threshold does not render maximal information, so the concurrence does not vanish. This happens because the scatterer and the electron are in an entangled state so the state of either of the two subsystems (electron or scatterer) is not completely determined (a mixed state). 

Beyond the concurrence minimum displayed by the dashed curve of Fig.\ref{fig3}, the two electron concurrence increases since one moves away from the resonance condition that determines the state of the scattered electron increasing the uncertainty of the electron state. The probability of an inelastic event is also reduced because the resonance condition is not met. 

The previous interpretation is the quantum information point of view for understanding tripartite entanglement (two fermions and a scatterer) that is treated analytically here by choosing to put all the involvement about the third party (scatterer) in the transmission and reflection amplitudes. From this point of view the concurrence minimum is associated directly with a peculiar $k$-dependent resonance where both the position and the width of the resonance depend on the energy of the incident particle. Using Eq.\ref{phase} and equating it to $-i$ (minus the imaginary unit) thus identifying the poles of the scattering amplitude we can determine both the position of the one particle resonance and its width\cite{Lipkin}. In the small $u_1$ inelastic coupling limit
\begin{equation}
E_R=\frac{\hbar^2}{2m}\left [\left ( \frac{u_0}{2}-\frac{u_1^2u_0}{u_0^2+4k^2}\right )\right  ]+\frac{i\hbar^2 k u_0 u_1^2}{4m[u_0^2+4k^2]},
\end{equation}
which shows a $k$ dependent position for the resonance and an asymmetric line shape, exactly the reverse of the behavior for the concurrence i.e. the maximum of the resonance is the minimum for the concurrence. Therefore the behavior of concurrence for the dashed curve in Fig.\ref{fig3} can be fully explained in terms of a bipartite entanglement scattering.

Finally, for the lowest excitation energy and both incoming particle momenta above the inelastic threshold $\sqrt{E_{exc}}$ (conditions satisfied for the whole $\Delta k$ region), the concurrence decays monotonically as $\Delta k$ increases. There are no thresholds for inelastic processes or transmission resonances involved, so the decay is monotone. For sufficiently large $\Delta k$, which when one of the momenta is fixed means that the second momentum is increasing, the transmission probability for such $k$ is increasing monotonously. As we go up in energy the delta barrier becomes more transparent, albeit inelastic, and the state of the second particle becomes better known reducing the concurrence. This reduction of the concurrence occurs for all the situations described above for sufficiently large incident  momentum.
 %-----------------------begin------------------Fig.4
\begin{figure}
\includegraphics[width=8cm]{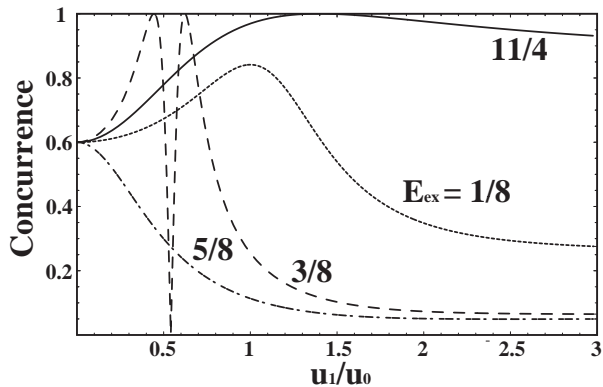}
\caption{\label{fig4} {The concurrence as a function of the inelastic coupling strength $g=u_1/u_0$. The values $k_1=u_0/2$ and $k_2=k_1+u_0$ have been fixed for the four indicated representative values of $(E_e-E_g)/(4E_{bind})$. The dotted curve corresponds to both momenta above the inelastic threshold. The dashed and dash-dotted curves correspond to one momentum above and the other below the threshold and the solid curve depicts purely elastic processes. Note the enhancement of the concurrence, for $u_1<3u_0/2$, as compared to the zero inelastic coupling limit. }}
\end{figure}
%----------------------end---------------------Fig.4

Figure \ref{fig4} depicts the concurrence as a function of the coupling to the excited state of the scatterer. In the figure we fix the incoming electron states (see figure caption) while the excitation energy, $E_{exc}$, takes on four representative values. The three situations correspond to i) dotted curve; both incoming states are above the inelastic threshold ii) dashed and dash-dot curve; one incoming state is above and the other below, and iii) solid curve; both states are below the threshold (purely elastic scattering). As a reference, for zero coupling to inelastic events the concurrence is the same for all curves. This can be shown to be general, given fixed values for $k_1$ and $k_2$, using an expansion for the concurrence at small coupling $u_1$  
\begin{eqnarray}
\label{concurrenceexpand}
\eta&\sim&\frac{2k_1k_2}{k_1^2+k_2^2}+\\
&&\frac{8k_1k_2(k_1^2-k_2^2)^2g^2}{(k_1^2+k_2^2)^2(4k_1^2+u^2_0-4u^2_0E_{exc})(4k_2^2
+u^2_0-4u_0^2E_{exc})},\nonumber
\end{eqnarray}
where the correction is valid when both momenta are above the inelastic threshold, such as the dotted curve in Fig.\ref{fig4}. The expression is completely symmetric between $k_1$ and $k_2$, as expected, and quadratically increases with $g$. Note that when the coupling is zero, the concurrence only depends on the incoming momenta. These are fixed in Fig.\ref{fig4} for all curves so they depart at the same value for zero coupling. At zero coupling no inelastic absorption can take place since there can be no energy transfer between the particle and the scatterer.

It is strikingly evident from Fig.\ref{fig4} that in all three ranges above that while stronger inelastic coupling reduces the concurrence, smaller couplings can actually enhance it beyond the limit of no coupling to excited states of the scatterer. In the inelastic limit for both scattered particles (case i) one can show this enhancement by looking at Eq.\ref{concurrenceexpand} valid for small inelastic coupling, where the correction is quadratic in the coupling $g=u_1/u_0$. As we saw in relation with Fig.\ref{fig3}, such non-monotonic features are related to the information on the electron states as we approach regular resonances or in regard to the approach to the entangled resonance at the inelastic threshold.
This behavior contrasts with voltage probe models\cite{Prada,Prada2} that yield a monotonic decrease of the entanglement degree as a function of the intensity of the coupling. Obviously the implicit mechanisms in such models that produce complete decorrelation of phase and/or energy destroy the details of the information exchanges that are explicit in our model. 

Another interesting feature of Fig.\ref{fig4} is that while for any fixed value of the excitation energy, the concurrence eventually decreases with the coupling, it does so to a plateau value beyond which no further reduction of the concurrence ensues. This means that we cannot make the scatterer absorb more than half of the input probability i.e. inelasticity cannot break unitarity in the electron subsystem beyond a certain point. Such upper limit for the probability of absorption can be straightforwardly derived from Eq.\ref{intensity}, where ${\rm Im}(\delta_0)\in(0,\infty)$, so that $|r|^2+|t|^2\geq 0.5$. Physically, it follows from the fact that when the ground and excited states of the scatterer are equally filled the coupling term in the Hamiltonian absorbs energy as strongly as it cedes it.

Looking at figures \ref{fig3} and \ref{fig4} with the eye for engineering the best point of operation for the entangler, one sees that one alternative is the purely elastic case for small $\Delta k$ avoiding regular resonances in the transmission or reflection. This situation is such that the post-selected electron states are maximally entangled. Nevertheless, in a real application, inelasticity or some other form of decoherence may be inevitable, so that one would have to find points of operation by modifying the inelastic coupling or changing the $\Delta k$ value to reach another optimal of operation. One such situation is $g=1$, $E_{ex}=1/8$ (see Fig.\ref{fig4}) and small $\Delta k$ (see Fig.\ref{fig3}).

Another point regarding practicality of our model is the validity of considering only a two level structure to  represent inelastic effects. In the arguments above, from the quantum information point of view, we have taken the route of asking for the knowledge about the states in the individual scattering event. We could, just as well, think of the problem as an ensemble of two level scatterers under a continuous stream of electrons, that can excite or elastically bound off the impurity . One thus establishes an ensemble average of individual entanglement generating events. Furthermore, even if a continuous stream of electrons hit a single impurity, the coupling to inelastic events depends on the presence of electrons at the site (through the delta at zero). A collision can both excite or relax the impurity so it will not saturate the system i.e. render the scatterers inactive or purely elastic. It is interesting that in a sequential electron scattering, a second electron can encounter an entangled system between the impurity and the previous electron collision. This brings about issues of time dependence which we have not addressed here and is left for future work.

The results obtained here also apply for the problem of electron-hole entanglement\cite{2}, which could have direct physical realizations by implementing partially reflecting barriers that are also absorbing. As we have shown, nontrivial mechanisms for entanglement in this case need only involve an energy absorbing event for the tunneling electron. In practice, coupling of the electron to phonons can provide the inelastic events. This has been studied in detail in the so called double barrier resonant tunneling devices, where electron tunneling is assisted by longitudinal optical phonons so that the electron can give up excess energy and tunnel resonantly\cite{Goldman}. This effect has recently been used as the basis of a coherent phonon emission device\cite{Makler}.

\section{summary and conclusions}
In this work we have analyzed the role of inelastic events, due to an excitable scatterer with an internal two level structure, in the entanglement production during a scattering process. The $S$ matrix formalism is adapted to treat the non-unitary case for two non-interacting particle scattering. The entangled state is post selected such that there is one particle reflected and one transmitted. We find the effects of two types of resonance structures, the first due to one particle perfect transmission or reflection in the elastic regime. In this regime the resonance causes a vanishing concurrence between the electrons as perfect knowledge of the one electron subspace is achieved. On the other hand, there is a second type of resonance we call entangled resonance, that occurs in the inelastic regime, where an entangled state is formed between one of the electrons and the scatterer. Such a resonance occurs at an energy dependent value and its width is asymmetric also energy dependent. Once more, as the resonance energy is approached, it gives away information of one of the electron state, but such information is limited by the mutual information between that electron and the scatterer.

A compelling feature produced by the inelastic entangled resonance is that the two particle concurrence can be enhanced by inelastic absorption beyond the elastic scattering result. This result is found both for one or both particles suffering energy loss to the excited scatterer. Such behavior contrasts with voltage probe models\cite{Prada,Prada2} that yield a monotonic decrease of the entanglement with the strength of the coupling. Of course, the voltage probe models for decoherence are phenomenologically concocted to produce complete decorrelation either in phase only or both in phase and energy. The more detailed exchange of information between scatterer and scattered entities is completely lost.  How to link both limits, that go from detailed quantum information exchanges to decorrelation in phase is currently unclear. 

Evidently, in the model considered here cannot be regarded as a bath in the regular sense, since the Hilbert space is possibly the smallest that can describe quantum inelasticity. One could say that in our case only one bit of information can be lost (this bit being the state of the impurity) whereas the number of bits lost is infinite in the case of a more realistic bath\cite{joos}. One interesting possibility to extend our analysis beyond this limitation is a decoherence mechanism described in reference \cite{FoaAntires}, where a coarse graining of the resonance/antiresonance structure conduce to the well known voltage probe results. This approach gathers increasing validity in the recent work of Kofler and Brukner\cite{Kofler}, where the energy coarse graining approach is justified on more general grounds.

\begin{acknowledgments}
The authors acknowledge support from FONACIT through Grants No. G-2001000712 and S3-2005000569.
 \end{acknowledgments}

\end{document}